\documentclass[pra,twocolumn]{revtex4-1}

\usepackage{graphicx}
\usepackage{amsmath}
 \usepackage{amssymb}
\usepackage{amsfonts}
\usepackage{upgreek}
\usepackage{comment}
\usepackage{color}
\usepackage{hyperref}

\begin{document}

\title{Self-organised Limit-Cycles, Chaos and Phase-Slippage\\ with a Superfluid inside an Optical Resonator}

\author{Francesco Piazza}
\author{Helmut Ritsch}

\affiliation{\small Institut f\"ur Theoretische Physik, Universit{\"a}t Innsbruck, A-6020~Innsbruck, Austria}

\begin{abstract}
We study dynamical phases of a driven Bose-Einstein condensate coupled to the light field of a high-$Q$ optical cavity. 
For high field seeking atoms at red detuning the system is known
to show a transition from a spatially homogeneous steady-state to a
self-organized regular lattice exhibiting super-radiant scattering into the cavity. For blue atom pump detuning the particles
are repelled from the maxima of the light-induced optical potential suppressing scattering.
We show that this generates a new dynamical instability of the self-ordered phase, leading to the appearance of self-ordered
stable limit-cycles characterized by large amplitude self-sustained
oscillations of both the condensate density and cavity field. The
limit-cycles evolve into chaotic behavior by period doubling. Large
amplitude oscillations of the condensate are accompanied by
phase-slippage through soliton nucleation at a rate which increases by
orders of magnitude in the chaotic regime. Different from a
superfluid in a closed setup, this driven dissipative superfluid is not destroyed by
the proliferation of solitons since kinetic energy is removed through
cavity losses.
\end{abstract}

\maketitle

\begin{figure}[htp]
\includegraphics[width=90mm]{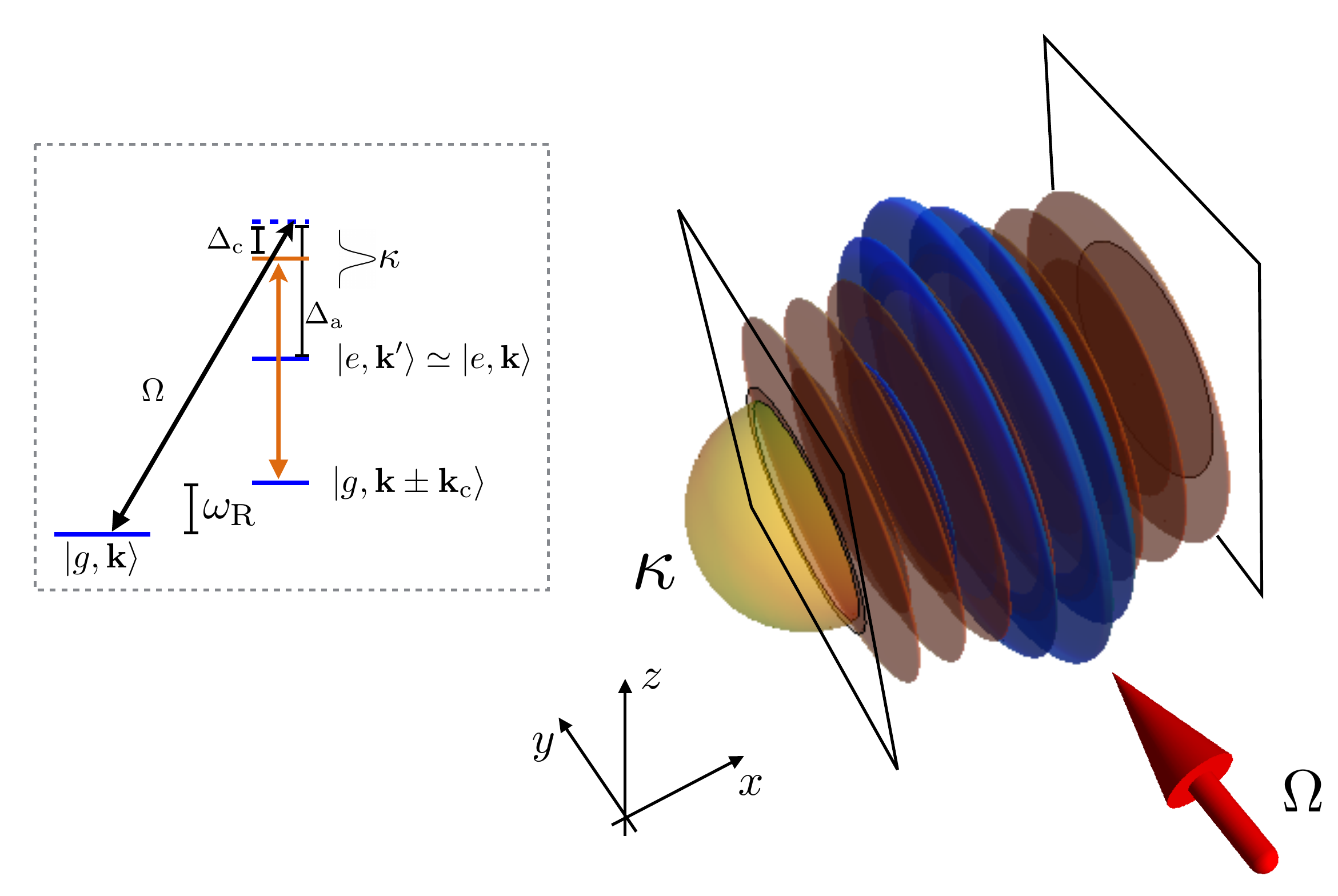}
\caption{A Bose-Einstein condensate of neutral atoms (blue
  surface) trapped inside an optical resonator is driven with a
  monochromatic laser with amplitude $\Omega$. The laser frequency is
  blue-detuned by an amount $\Delta_{\rm a}$ with respect to an
  internal atomic transition $g\leftrightarrow e$ and by $\Delta_c$
  with respect a standing-wave cavity mode
  $\sim\cos(k_{\rm c}x)$ defining the characteristic recoil frequency
  $\omega_{\rm R}=\hbar k_{\rm c}/2m$ with atomic mass $m$. 
  The coupling between a single atom and the cavity mode has a
  strength $g_0$. For large $\Delta_{\rm a}\gg (\Delta_{\rm c},\kappa,
  g_0, \omega_{\rm R})$, with $\kappa$ being the cavity linewidth due to
  leakage out of the mirrors, the dispersive regime is reached, where
  the atoms stay effectively in their internal ground state. They experience negligible spontaneous emission and
  an optical potential (orange surface) arising from the interfering
  cavity and pump fields (see Eq.\eqref{eq:gp}). For simplicity we assume a sufficiently strong
  extra atom trapping in the radial $y,z$ directions to confine the dynamics onto the cavity axis $x$ and generate an effective 1D geometry.}
\label{fig:setup}
\end{figure}

\section{Introduction}

The experimental realization of strong collective coupling between a large number of (ultra)cold atoms and the
electromagnetic field of Fabry-Perot cavities
\cite{vuletic_2003,courteille_2007,reichel_2007,st_kurn_2007,eth_2010,barrett_2012,hemmerich_SR_2014,lev_multimode_exp_2015},
nanophotonic fibers
\cite{rauschenb_2010,kimble_2012_fiber,lukin_2013_fibercavity} or photonic crystals \cite{kimble_2014_crystal}
opens up new interesting routes both in the field of quantum optics and condensed matter physics. 
From the latter point of view, the interesting new ingredient is provided by very well controllable strong long-range photon-mediated atom-atom
interactions appearing due to the back-action of even a single atom onto the light field. Strong long-range interactions can
indeed lead to several intriguing phenomena \cite{ruffo_review,morigi_pret}
and are crucial in many intensively explored condensed matter
phases, like supersolids \cite{batrouni_supersolid_1995} or
topological states \cite{zoller_topological_2006}. 
What is more, differently from typical condensed matter situations,
these light-mediated interactions are in general i)
retarded, since the photon field owns intrinsic timescales which can
be made comparable with atomic scales, ii) non-conservative, since the
system is typically driven and dissipates energy through atomic
spontaneous emission and photon losses through the cavity mirrors.

A striking consequence of cavity field mediated interactions is the appearance of self-ordered phases \cite{ritsch_2002,cavity_rmp,chang_2013,griesser_2013,keeling_2014,piazza_fermi,zhai_2014,bonifacio_freespace_so_2015}
where the particles break a translation symmetry by forming a spatial
pattern determined by the characteristic interaction length scale. 
This phenomenon has been observed experimentally both for a thermal
gas \cite{vuletic_2003,barrett_2012} and an ultracold Bose-Einstein
condensate (BEC) \cite{eth_2010,hemmerich_SR_2014} coupled to a
standing-wave mode of an optical cavity as sketched in
Fig.~\ref{fig:setup}. In the regimes considered so far, a thermal gas
and a BEC  share the same qualitativ behavior \cite{piazza_bose}.
Interestingly the self-ordering of a BEC can be closely mapped to the superradiant transition of the famous Dicke model of N two-state atoms coupled to a single cavity mode \cite{dicke_54,eth_2010}.
If fundamental collective mechanical excitations of the BEC are effectively treated as quantum mechanical oscillators, one also gets a simulator for a general
optomechanics setup close to zero temperature \cite{stamper2014cavity,nagy_2010}.

Here we study a new regime of quantum gas cavity QED \cite{mekhov2012quantum}, where the 
self-ordering phase transition is tight to dynamical instabilities. This gives rise to new types of nonequilibrium phases
for which the peculiarities of the BEC play a crucial role. This novel behaviour appears upon a rather innocent looking
change of operating conditions, namely by choosing the frequency of
the driving laser larger than the atomic internal transition frequency
(blue detuning). In this regime the atoms are low field-seekers as
opposed to high field seekers in the typically considered red detuning
case (see Fig.~\ref{fig:setup}).  Naively, one would expect that this
prevents any self-ordering as the atoms are pushed towards field
minima, where light scattering is suppressed. Surprisingly, a closer
look reveals that the complex interplay of collective coherent
scattering and optical dipole forces still can generate a self-ordered
phase at sufficient pump strength. However, the particles are now
localized at cavity field nodes and this order gets dynamically
unstable again at only a somewhat higher critical pump intensity, as illustrated in Fig.~\ref{fig:phase_diagram}.

Interestingly, this instability does not simply lead to heating and disintegration of the order, but we find the emergence of limit-cycles, whereby the condensate performs large periodic self-sustained oscillations between different ordered patterns. The atomic density oscillations are tightly coupled to the oscillation of the cavity field with the same frequency, as shown in Fig.~\ref{fig:phase_slips}A). This provides a build-in non-destructive monitoring tool of the nonlinear dynamics. 
By further increase of  the drive strength the limit-cycles turn into
chaotic dynamics by doubling their period
(see Fig.~\ref{fig:chaos}).

Dynamical instabilities toward limit-cycles evolving into chaos
have been also observed with nanomechanical oscillators coupled to light
\cite{carmon_opto_exp_2005,carmon_opto_chaos_exp_2007,marquardt_exp_2008}. Limit-cycles
have been studied within the open Dicke model as well \cite{simons_2010}, where chaos
appears in the closed-system limit \cite{emary03,haake_chaos_dicke_2012,hirsch_dicke_chaos_2015}. 
Self-sustained oscillations of a BEC inside a driven cavity have been
observed \cite{eth_2007} and shown to be very well described through an optomechanical model. 
In the same setup, a transition to chaos has also been predicted
\cite{oppo_bec_lc_2014}, analogous to the one appearing with nonlinear
dielectric medium \cite{ikeda_1980}. Recently, superfluid Josephson
dynamics of a BEC coupled to a driven cavity have also been
theoretically studied \cite{mazzarella_jos_cav_2015}.

In this work, we show how the dynamical self-ordered regimes of large
nonlinear excitations let
the BEC hallmarks emerge clearly, which prevents an understanding of the
system via (generalized) optomechanical models. In particular, once direct short-range
atom-atom interactions are taken into account, the superfluid nature
of the BEC manifests itself through the onset of phase-slippage
dynamics
\cite{anderson_rmp_1966,avenel_ps_1988,packard_ps_1992,nist_ps_2013}, whereby the condensate lowers the kinetic energy stored in
the phase of the macroscopic wave function by creating phase
singularities in the form of nonlinear dispersive waves (solitons in
our one-dimensional model). While phase-slips take place periodically
in the lymit-cycle phase, they appear irregularly and at a much faster
rate in the chaotic regime, as illustrated in
Fig.~\ref{fig:phase_slips}. Interestingly, different from a
superfluid in a closed system where phase-slip proliferation
eventually destroys the superfluid, here the dissipation through
cavity losses counteracts this heating process by subtracting energy from the system.

Our findings introduce a new scenario where nonlinear chaotic dynamics,
self-organisation and superfluidity appear together in a driven/dissipative system, 
bridging between the optomechanics/nonlinear optics and condensed matter communities.

\section{Dynamical phase diagram}

\begin{figure}[htp]
\vspace{5mm}
\includegraphics[width=90mm]{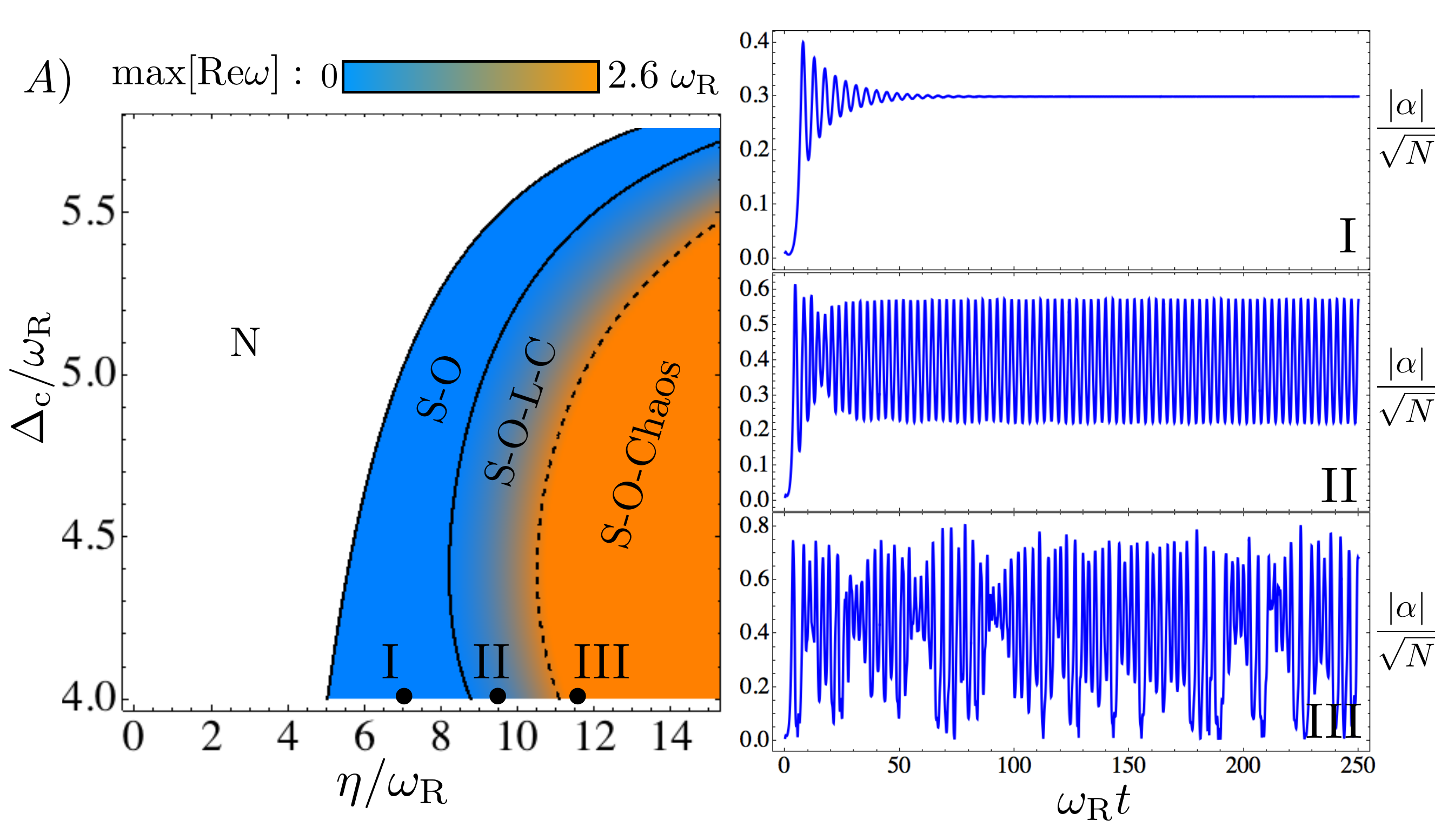}
\caption{ A) Nonequilibrium phase diagram for the driven Bose-Einstein condensate as in
  the setup of Fig.~\ref{fig:setup} as a function of the effective
  pump strength $\eta=\sqrt{N}g_0\Omega/\Delta_{\rm a}$ and the detuning
  $\Delta_{\rm c}$, for $\kappa=10\omega_{\rm R}$, $U_0
  N=12.1\omega_{\rm R}$, and $g_{\rm aa}=0$. The phase is determined by the long-time behavior
  of the system starting from an empty cavity and a spatially
  homogenous condensate along $x$. The color scale indicates the
  growth rate $\mathrm{Re}\omega$ of the most unstable collective excitation mode above the
  steady-state. The corresponding time evolution of
  the cavity mode amplitude $\alpha$ is shown in panels I,II, and III. Four different phases emerge. 
  In the normal (N) phase, the steady-state is still an empty cavity
  with a homogeneous condensate. In the self-organised (S-O) phase, the
  steady state is stable and has a finite $\alpha$ accompanied by the
  corresponding condensate density modulation in space, optimising the light scattering
  into the cavity. In the self-organised limit-cycle (S-O-L-C) phase,
  the steady state is unstable and evolves into periodic
  self-sustained oscillations of large amplitude about a finite value
  of $\alpha$. At every time during the oscillation the condensate is
  self-organised, i.e. has chosen the density modulation giving rise
  to the instantaneous cavity amplitude. In the self-organised chaotic
  (S-O-Chaos) phase, the collective oscillations loose their
  periodicity. The transition from limit-cycles to chaos takes place
  by period doubling, as illustrated in Fig.~\ref{fig:chaos}.  
}
\label{fig:phase_diagram}
\end{figure}

As sketched in Fig.~\ref{fig:setup}, we consider an ultracold atomic gas trapped along the axis a single mode of a Fabry-Perot
cavity. The gas is illuminated with coherent light which is blue detuned to an atomic transition. Corresponding experiments with BECs of ultracold atoms
\cite{eth_2010,hemmerich_SR_2014} have shown to be well described by coupled classical field equations \cite{cavity_rmp} describing
the mean-field dynamics of the BEC with the Gross-Pitaevskii equation
\cite{string_pit}
\begin{align}
\label{eq:gp}
i\hbar\partial_t\psi(x,t)&=\left[-\frac{\hbar^2\partial^2_{xx}}{2m}+g_{\rm
    aa}|\psi(x,t)|^2\!\!+\hbar U_0|\alpha(t)|^2\!\!\cos^2(k_{\rm c}x)\right.\nonumber\\
&\left.+2\hbar(\eta/\sqrt{N})\mathrm{Re}[\alpha(t)]\cos(k_{\rm c}x)\right.\bigg]\psi(x,t)
\end{align}
and the cavity field by its coherent component dynamics $\alpha(t)$  :
\begin{align}
\label{eq:field_dyn}
i\partial_t\alpha(t)=&\left[-\Delta_{\rm c}-i\kappa+U_0\int\! dx |\psi(x,t)|^2\cos^2(k_{\rm c}x)\right]\alpha(t)\nonumber\\
&+(\eta/\sqrt{N})\int\! dx |\psi(x,t)|^2\cos(k_{\rm c}x)\;.
\end{align}
In the Gross-Pitaevskii framework, all $N$ particles in the condensate,
i.e. occupying the same quantum state, share the same BEC wavefunction
$\psi(x,t)$ normalized to $N$.
The first term in Eq.~\eqref{eq:gp} describes the kinetic energy of
the quantum motion of the particles with mass $m$, which we restricted along
the cavity axis $x$ upon assuming additional trapping in the other
directions. This can be easly achieved in ultracold atom
experiments \cite{zwerger_rmp}. Correspondingly, the direct atom-atom interaction
strength $g_{\rm aa}$ is the effective coupling for the
one-dimensional problem \cite{string_pit}. $U_0=g_0^2/\Delta_{\rm a}$
is the potential depth per photon felt by an atom (see
Eq.~\eqref{eq:gp}) as well as the energy shift of the cavity resonance
per atom (see Eq.~\eqref{eq:gp}). It results from emission and
absorption of a photon from and into the cavity mode and is therefore
$\propto\cos^2(k_{\rm c}x)$ in the dispersive regime where the excited
atomic state can be adiabatically
eliminated. The further terms containing $\eta/\sqrt{N}=g_0\Omega/\Delta_{\rm
  a}$ result from absorption from the pump and emission into
the cavity mode (or viceversa) and introduce a further optical
potential $\propto\cos(k_{\rm c}x)$ for the atoms and an effective
pump term for the cavity field. Finally, the loss of photons through
the cavity mirrors is reflected by the field damping term $-i\kappa$. 
Eqs.~\eqref{eq:gp},\eqref{eq:field_dyn} describe the dynamical
back-action between atoms and light, since the optical potential felt by the atoms is determined by
$\alpha(t)$, whose dynamics in turn depends on the atom wave
function. Equivalently, the field $\alpha$ mediates infinitely
long-range and retarded interactions between the BEC atoms.

The phase diagram of Fig.~\ref{fig:phase_diagram} is obtained by
solving Eqs.~\eqref{eq:gp},\eqref{eq:field_dyn} and analysing their
long-time behavior. As initial conditions we choose a homogeneous BEC:
$\psi(x,0)=\sqrt{n}$ with $n=N/L$ the system's density in one
dimension and
an infinitesimally occupied cavity $\alpha(0)\ll 1$, which is needed
as a seed since Eqs.~\eqref{eq:gp},\eqref{eq:field_dyn} don't include
noise.
A transition toward stable self-ordering in observed at a critical
pump strength \cite{domokos_2008} $\hbar^2\eta_{\rm
  crit}^2=(\hbar\omega_{\rm R}+2g_{\rm
  aa}n)(\delta_{\rm c}+\kappa^2/\delta_{\rm c})/2$, with the dispersively shifted
cavity detuning $\delta_{\rm c}=-\Delta_{\rm c}+U_0N/2$. This
transition is due to the fact that above a critical pump
strength the homogeneous condensate is unstable with respect to density
modulations at the cavity wavelength $\lambda_{\rm c}$. The latter
indeed tend to optimise light scattering into the cavity which in
turn enhances the density modulation in a runaway process. This is
stabilized by losses, providing convergence toward a steady state with
finite cavity amplitude $\alpha\neq 0$ and density modulation. This happens both for high ($U_0<0$) and low
($U_0>0$) field-seeking atoms. However, as we show here, in the latter
case the stable self-ordered steady state exists only up to a second
critical pump strength, above which it becomes dynamically unstable
and no time independent steady state can be found anymore. This originates from the
competition between the cavity $\propto\cos^2(k_{\rm c}x)$ generated potential pushing the particles to the cavity field nodes
where scattering is suppressed and the pump-cavity interference component 
$\propto\cos(k_{\rm c}x)$ pushing the particles to competing wavelength order.

\section{Self-ordered limit-cycles and chaos}
\begin{figure*}[htp]
\vspace{5mm}
\includegraphics[width=180mm,height=110mm]{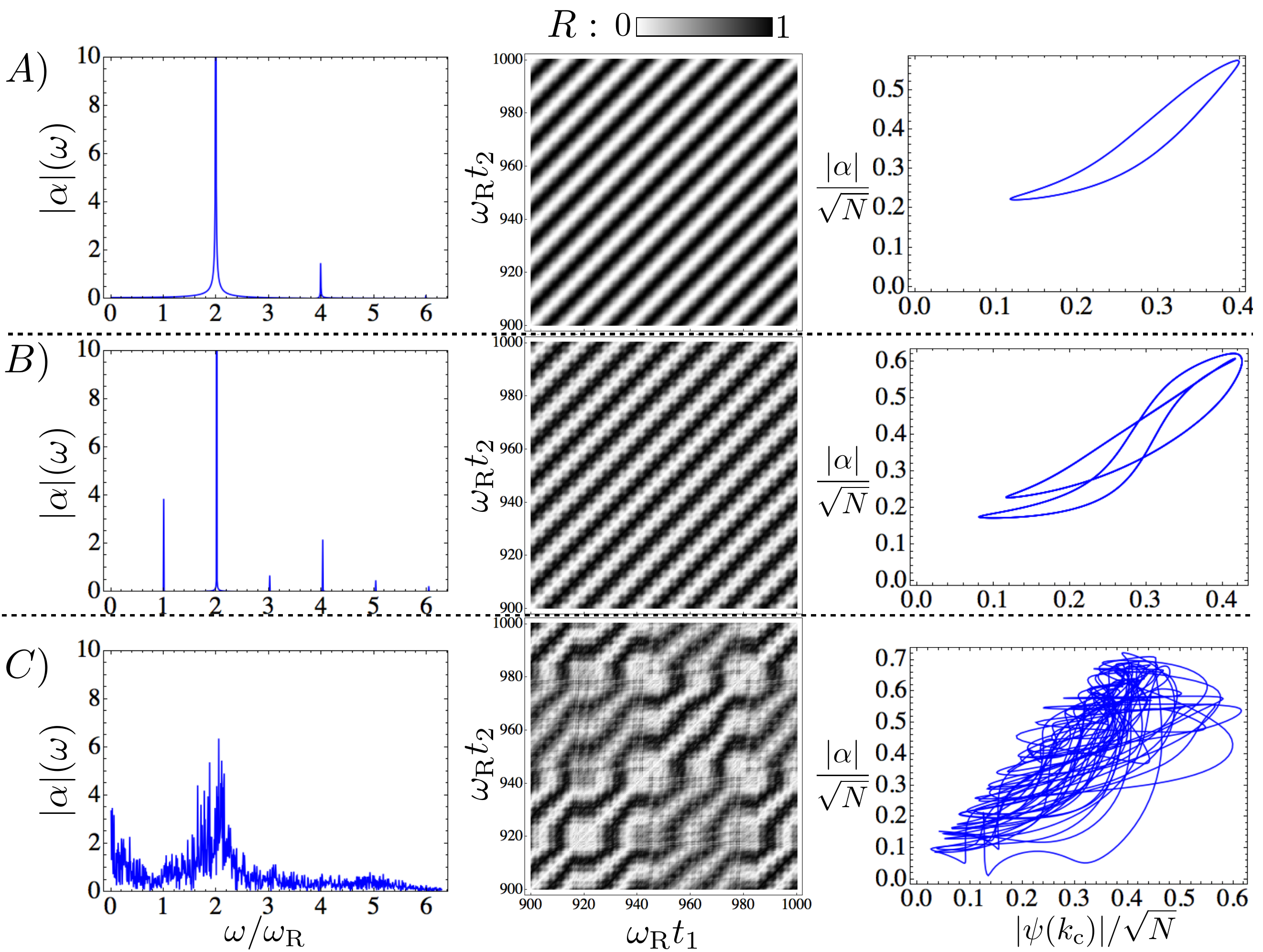}
\caption{Transition from limit-cycles to chaos. The Fourier transform
  of the cavity amplitude time-oscillations (left column) is shown
  together with the condensate wave function recurrence \cite{marwan_2007}
  $R(t_1,t_2)=\int dx|\psi(x,t_1)-\psi(x,t_2)|^2$ (middle column) and
  phase-space trajectories in the $|\alpha|-|\psi(k_{\rm c})|$ plane
  (right column). Here $\psi(k_{\rm c})$ is the Fourier component of
  the condensate wavefunction calculated at the cavity wave
  number $k_{\rm c}$. For the same parameters as in
  Fig.~\ref{fig:phase_diagram}, results are shown for three different
  pump strengths: A) $\eta=9.5\omega_{\rm R}$, B)
  $\eta=10\omega_{\rm R}$, and C) $\eta=11\omega_{\rm R}$ at
  $\Delta_{\rm c}=4\omega_{\rm R}$.
  The spectra clearly show period doubling (between A) and B)) before
  the onset of the noisy background in C). This indication for the
  appearance of chaos is confirmed by the large-scale structures in
  the recurrence $R$, as opposed to the stripes characterising a
  periodic behavior, visible in A) and B). In the latter, period
  doubling shows up as further small-scale structures perpendicular to
  the stripes in $R$. Phase-space trajectories confirm this picture,
  whereby the chaotic behavior is signalled by the absence of sharp
  closed trajectories.
}
\label{fig:chaos}
\end{figure*}
\begin{figure*}[htp]
\vspace{5mm}
\includegraphics[width=180mm]{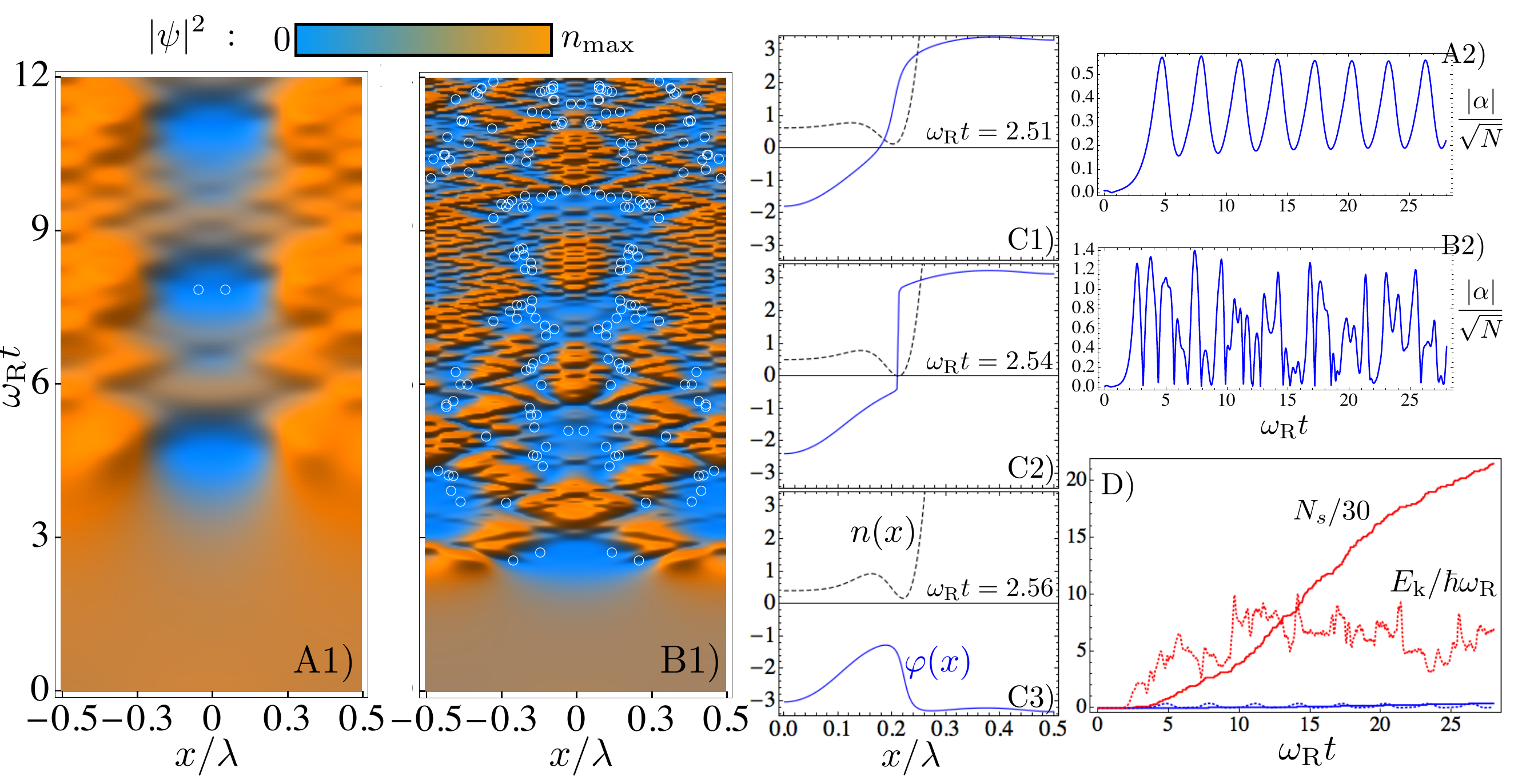}
\caption{Phase-slippage through soliton nucleation for a finite value of the atom-atom interaction $g_{\rm
    aa}=1.0\hbar\omega_{\rm R}$. Panels A) and B)
  show the condensate density (color-scale) as a function of position
  and time (A1 and B1) as well as the time-evolution of the cavity
  amplitude (A2 and B2) for $\eta=12.0\omega_{\rm R}$, which corresponds to a limit-cycle
  dynamics (A) and $\eta=22.0\omega_{\rm R}$, which corresponds to chaotic
  dynamics (B). In A),B), white circles indicate the space-time
  coordinate of phase-slips, i.e. phase singularities in the form of
  solitons having a zero-density core together with a $\pi$ phase
  difference across it. The dynamics of a single phase-slip is
  illustrated in panels C1) to C3) for $\eta=22.0\omega_{\rm
    R}$. The phase singularity is present in C2) as a $\pi$-phase jump
  localized at the dark ($n(x)=0$) soliton position. The latter has a
  size of the order of the healing length $\xi=1/\sqrt{2mg_cn}\sim
  0.1\lambda$. 
  The phase-slip consists in changing the sign of the phase jump $\pi\to -\pi$ (they
  are equivalent at the point $n(x)=0$). Subsequently, in C3) the soliton
  increases its minimum density from zero and moves away, carrying the
  energy subtracted from the initial phase gradient in C1).
  Panel D) shows the total number of phase-slips $N_s(t)$ (solid line) generated up
  time $t$ together with the
  condensate kinetic energy $E_{\rm k}$ (dotted line) as a
  function of time, for $\eta=12.0\omega_{\rm R}$ (in blue) and
$\eta=22.0\omega_{\rm R}$ (in red). While in the limit-cycle
regime phase-slips are generated following the slow periodic
oscillations (two phase-slips every second period in the case shown), the rate of slippage in the chaotic
regime is orders of magnitude faster. Still, soliton proliferation is
counteracted by dissipation through cavity losses so that in the
chaotic regime the kinetic energy intially grows but then flucuates
around a constant value.
}
\label{fig:phase_slips}
\end{figure*}

Above the second critical driving strength the system resolves the above
competion by performing self-sustained oscillations (limit-cycles)
between self-ordered configurations of the condensate (see
Fig.~\ref{fig:phase_slips}) A1) and cavity
field, shown in Fig.~\ref{fig:phase_diagram} II, \ref{fig:chaos}
A)-B), and \ref{fig:phase_slips} A2). 

As illustrated in Fig.~\ref{fig:chaos}, the limit-cycles involve several well defined frequency components,
whose number is doubled before the system dynamics turns eventually
chaotic. The onset of chaos by period doubling is a well known
phenomenon in nonlinear systems and has been observed with
nanomechanical oscillators coupled to light \cite{carmon_opto_chaos_exp_2007}.
Here, like in these systems, chaos develops at zero temperature and without external periodic
modulations or (delayed) feedback control. The characteristic
timescale is given by the recoil frequency $\omega_{\rm R}$ which is
an intrinsic property of the BEC-cavity system.

Different from what so far studied with nanomechanical media,
nonlinear dielectrics and non-directly-driven BECs \cite{eth_2007,oppo_bec_lc_2014}, the oscillations
found here have a large (non-perturbative) amplitude where the light intensity scales with $N$
due to the self-ordering of the atoms. The latter generates
non-trivial time-dependent spatial structures.

\section{Driven-dissipative phase-slippage}

While spatial structures with a length scale set by the cavity
wavelength $\lambda_c$ would be present had we considered any
(e.g. thermal) ensemble of driven polarisable particles, the peculiar
nature of their size and dynamics that we oberve here originates from the combination of
macroscopic phase coherence (encoded in the BEC wavefunction) together
with short-range atom-atom interactions (inducing the term $g_{\rm
  aa}|\psi|^2$ term in Eq.~\eqref{eq:gp}). These two properties in
turn are at the core of the superfluid behavior of the BEC. They indeed
provide the condensate with a finite ``phase-rigidity'', i.e. a
finite energy cost $\propto \int\! dx (\partial_x\varphi)^2$ of creating a phase gradient in the wave function
$\psi=|\psi|\exp(i\varphi)$, which is associated with a finite
flow velocity of the superfluid $v=\hbar\partial_x\varphi/m$
\cite{leggett2006quantum}. 

As illustrated in Fig.~\ref{fig:phase_slips}, the driven BEC coupled
to the lossy cavity, once entering the nonlinear oscillating regime,
shows large phase gradients and thus accumulates the corresponding energy.
The most efficient way for the superfluid to get rid
of this extra energy is to convert it into nonlinear dispersive waves
through the process of phase-slippage \cite{anderson_rmp_1966}. In one dimension these waves
are solitons, for which phase-slips take place as described in
Fig.~\ref{fig:phase_slips} C). 
We observe indeed phase-slippage as the system enters the dynamically
unstable regime. During the limit-cycles, phase slips take
place periodically and are synchronised with the oscillations. The
rate of slippage is slow (see Fig.~\ref{fig:phase_slips} D)) and the
solitons are nucleated always at the same position, shown in
Fig.~\ref{fig:phase_slips} A1). On the contrary, once entered the
chaotic regime, phase-slippage takes place at an orders of magnitude
faster rate and the nucleation of solitons becomes irregular in space
and time. 

Interestingly, this fast proliferation of solitons does not eventually
destroy the BEC as it usually happens in an isolated
superfluid. In the latter case the large number of solitons causes the
phase of the wavefunction to change fast and over short lenghscales, thereby destroying phase
coherence. In addition, solitons contain a large kinetic energy which can be converted into the thermal
component. However, in our driven/dissipative system the proliferation of solitons is counteracted
by light scattering processes transferring energy into the cavity
field and expelling it through photon losses. Indeed, we oberve that
even though solitons are produced at a fast rate, the BEC kinetic energy
$E_{\rm k}=\int\! dx\hbar^2|\partial_x\psi|^2/2m$, after an initial increase,
fluctuates around a finite value of the order of the recoil
energy $\hbar\omega_{\rm R}$, as shown in Fig.~\ref{fig:phase_slips}.

\section{Conclusions and outlook}

Surprisingly, dynamical self-ordering of a BEC inside a Fabry-Perot cavity appears also in the blue detuned regime but with decisive new properties.
While collective scattering is enhanced by ordering it is inhibited by repulsion from field antinodes. This competing effects
generate self-sustained large amplitude nonlinear oscillations, which reconcile ordering and superradiant scattering averaged over an oscillation cycle.
For stronger pump the oscillations exhibit period doubling and evolve into chaotic behavior. This complex dynamics takes place
without any external modulation or feedback and is an intrinsic property of the driven dissipative nature of the dynamics.
The nonlinear atom-field dynamics here is accompanied by
phase-slippage, which comes about due to the superfluid nature of the
interacting BEC. In one dimension phase-slippage nucleates solitons whose proliferation in the chaotic regime is counteracted by cavity cooling.

In two spatial dimensions vortex nucleation is instead expected. It
would be interesting to study how their proliferation and dynamics is affected by light
scattering into the lossy cavity.
In this spirit, the extension of the present study to $d=2$ should open up
new directions in the field of quantum turbulence \cite{barenghi_review_2014} in
driven/dissipative condensates \cite{sieberer_2014}. 

We finally point out that the system studied here is already available experimentally since it
involves a change from red- to blue-detuning of the driving laser with
respect to the atomic resonance in the setups used in the ETH \cite{eth_2010} and Hamburg
\cite{hemmerich_SR_2014} laboratories.

\begin{acknowledgments}
We thank Philipp Strack for useful comments on the manuscript.
FP acknowledges support by the APART fellowship of the Austrian
Academy of Sciences. HR is supported by the Austrian Science Fund
project I1697-N27.
\end{acknowledgments}

\bibliographystyle{aipnum4-1}
\bibliography{mybib}

\begin{thebibliography}{51}%
\makeatletter
\providecommand \@ifxundefined [1]{%
 \@ifx{#1\undefined}
}%
\providecommand \@ifnum [1]{%
 \ifnum #1\expandafter \@firstoftwo
 \else \expandafter \@secondoftwo
 \fi
}%
\providecommand \@ifx [1]{%
 \ifx #1\expandafter \@firstoftwo
 \else \expandafter \@secondoftwo
 \fi
}%
\providecommand \natexlab [1]{#1}%
\providecommand \enquote  [1]{``#1''}%
\providecommand \bibnamefont  [1]{#1}%
\providecommand \bibfnamefont [1]{#1}%
\providecommand \citenamefont [1]{#1}%
\providecommand \href@noop [0]{\@secondoftwo}%
\providecommand \href [0]{\begingroup \@sanitize@url \@href}%
\providecommand \@href[1]{\@@startlink{#1}\@@href}%
\providecommand \@@href[1]{\endgroup#1\@@endlink}%
\providecommand \@sanitize@url [0]{\catcode `\\12\catcode `\$12\catcode
  `\&12\catcode `\#12\catcode `\^12\catcode `\_12\catcode `\%12\relax}%
\providecommand \@@startlink[1]{}%
\providecommand \@@endlink[0]{}%
\providecommand \url  [0]{\begingroup\@sanitize@url \@url }%
\providecommand \@url [1]{\endgroup\@href {#1}{\urlprefix }}%
\providecommand \urlprefix  [0]{URL }%
\providecommand \Eprint [0]{\href }%
\providecommand \doibase [0]{http://dx.doi.org/}%
\providecommand \selectlanguage [0]{\@gobble}%
\providecommand \bibinfo  [0]{\@secondoftwo}%
\providecommand \bibfield  [0]{\@secondoftwo}%
\providecommand \translation [1]{[#1]}%
\providecommand \BibitemOpen [0]{}%
\providecommand \bibitemStop [0]{}%
\providecommand \bibitemNoStop [0]{.\EOS\space}%
\providecommand \EOS [0]{\spacefactor3000\relax}%
\providecommand \BibitemShut  [1]{\csname bibitem#1\endcsname}%
\let\auto@bib@innerbib\@empty
\bibitem [{\citenamefont {Black}, \citenamefont {Chan},\ and\ \citenamefont
  {Vuleti\ifmmode~\acute{c}\else \'{c}\fi{}}(2003)}]{vuletic_2003}%
  \BibitemOpen
  \bibfield  {author} {\bibinfo {author} {\bibfnamefont {A.~T.}\ \bibnamefont
  {Black}}, \bibinfo {author} {\bibfnamefont {H.~W.}\ \bibnamefont {Chan}}, \
  and\ \bibinfo {author} {\bibfnamefont {V.}~\bibnamefont
  {Vuleti\ifmmode~\acute{c}\else \'{c}\fi{}}},\ }\href {\doibase
  10.1103/PhysRevLett.91.203001} {\bibfield  {journal} {\bibinfo  {journal}
  {Phys. Rev. Lett.}\ }\textbf {\bibinfo {volume} {91}},\ \bibinfo {pages}
  {203001} (\bibinfo {year} {2003})}\BibitemShut {NoStop}%
\bibitem [{\citenamefont {Slama}\ \emph {et~al.}(2007)\citenamefont {Slama},
  \citenamefont {Bux}, \citenamefont {Krenz}, \citenamefont {Zimmermann},\ and\
  \citenamefont {Courteille}}]{courteille_2007}%
  \BibitemOpen
  \bibfield  {author} {\bibinfo {author} {\bibfnamefont {S.}~\bibnamefont
  {Slama}}, \bibinfo {author} {\bibfnamefont {S.}~\bibnamefont {Bux}}, \bibinfo
  {author} {\bibfnamefont {G.}~\bibnamefont {Krenz}}, \bibinfo {author}
  {\bibfnamefont {C.}~\bibnamefont {Zimmermann}}, \ and\ \bibinfo {author}
  {\bibfnamefont {P.~W.}\ \bibnamefont {Courteille}},\ }\href {\doibase
  10.1103/PhysRevLett.98.053603} {\bibfield  {journal} {\bibinfo  {journal}
  {Phys. Rev. Lett.}\ }\textbf {\bibinfo {volume} {98}},\ \bibinfo {pages}
  {053603} (\bibinfo {year} {2007})}\BibitemShut {NoStop}%
\bibitem [{\citenamefont {Colombe}\ \emph {et~al.}(2007)\citenamefont
  {Colombe}, \citenamefont {Steinmetz}, \citenamefont {Dubois}, \citenamefont
  {Linke}, \citenamefont {Hunger},\ and\ \citenamefont
  {Reichel}}]{reichel_2007}%
  \BibitemOpen
  \bibfield  {author} {\bibinfo {author} {\bibfnamefont {Y.}~\bibnamefont
  {Colombe}}, \bibinfo {author} {\bibfnamefont {T.}~\bibnamefont {Steinmetz}},
  \bibinfo {author} {\bibfnamefont {G.}~\bibnamefont {Dubois}}, \bibinfo
  {author} {\bibfnamefont {F.}~\bibnamefont {Linke}}, \bibinfo {author}
  {\bibfnamefont {D.}~\bibnamefont {Hunger}}, \ and\ \bibinfo {author}
  {\bibfnamefont {J.}~\bibnamefont {Reichel}},\ }\href@noop {} {\bibfield
  {journal} {\bibinfo  {journal} {Nature}\ }\textbf {\bibinfo {volume} {450}},\
  \bibinfo {pages} {272} (\bibinfo {year} {2007})}\BibitemShut {NoStop}%
\bibitem [{\citenamefont {Gupta}\ \emph {et~al.}(2007)\citenamefont {Gupta},
  \citenamefont {Moore}, \citenamefont {Murch},\ and\ \citenamefont
  {Stamper-Kurn}}]{st_kurn_2007}%
  \BibitemOpen
  \bibfield  {author} {\bibinfo {author} {\bibfnamefont {S.}~\bibnamefont
  {Gupta}}, \bibinfo {author} {\bibfnamefont {K.~L.}\ \bibnamefont {Moore}},
  \bibinfo {author} {\bibfnamefont {K.~W.}\ \bibnamefont {Murch}}, \ and\
  \bibinfo {author} {\bibfnamefont {D.~M.}\ \bibnamefont {Stamper-Kurn}},\
  }\href {\doibase 10.1103/PhysRevLett.99.213601} {\bibfield  {journal}
  {\bibinfo  {journal} {Phys. Rev. Lett.}\ }\textbf {\bibinfo {volume} {99}},\
  \bibinfo {pages} {213601} (\bibinfo {year} {2007})}\BibitemShut {NoStop}%
\bibitem [{\citenamefont {Baumann}\ \emph {et~al.}(2010)\citenamefont
  {Baumann}, \citenamefont {Guerlin}, \citenamefont {Brennecke},\ and\
  \citenamefont {Esslinger}}]{eth_2010}%
  \BibitemOpen
  \bibfield  {author} {\bibinfo {author} {\bibfnamefont {K.}~\bibnamefont
  {Baumann}}, \bibinfo {author} {\bibfnamefont {C.}~\bibnamefont {Guerlin}},
  \bibinfo {author} {\bibfnamefont {F.}~\bibnamefont {Brennecke}}, \ and\
  \bibinfo {author} {\bibfnamefont {T.}~\bibnamefont {Esslinger}},\ }\href@noop
  {} {\bibfield  {journal} {\bibinfo  {journal} {Nature}\ }\textbf {\bibinfo
  {volume} {464}},\ \bibinfo {pages} {1301} (\bibinfo {year}
  {2010})}\BibitemShut {NoStop}%
\bibitem [{\citenamefont {Arnold}, \citenamefont {Baden},\ and\ \citenamefont
  {Barrett}(2012)}]{barrett_2012}%
  \BibitemOpen
  \bibfield  {author} {\bibinfo {author} {\bibfnamefont {K.~J.}\ \bibnamefont
  {Arnold}}, \bibinfo {author} {\bibfnamefont {M.~P.}\ \bibnamefont {Baden}}, \
  and\ \bibinfo {author} {\bibfnamefont {M.~D.}\ \bibnamefont {Barrett}},\
  }\href {\doibase 10.1103/PhysRevLett.109.153002} {\bibfield  {journal}
  {\bibinfo  {journal} {Phys. Rev. Lett.}\ }\textbf {\bibinfo {volume} {109}},\
  \bibinfo {pages} {153002} (\bibinfo {year} {2012})}\BibitemShut {NoStop}%
\bibitem [{\citenamefont {Ke\ss{}ler}\ \emph {et~al.}(2014)\citenamefont
  {Ke\ss{}ler}, \citenamefont {Klinder}, \citenamefont {Wolke},\ and\
  \citenamefont {Hemmerich}}]{hemmerich_SR_2014}%
  \BibitemOpen
  \bibfield  {author} {\bibinfo {author} {\bibfnamefont {H.}~\bibnamefont
  {Ke\ss{}ler}}, \bibinfo {author} {\bibfnamefont {J.}~\bibnamefont {Klinder}},
  \bibinfo {author} {\bibfnamefont {M.}~\bibnamefont {Wolke}}, \ and\ \bibinfo
  {author} {\bibfnamefont {A.}~\bibnamefont {Hemmerich}},\ }\href {\doibase
  10.1103/PhysRevLett.113.070404} {\bibfield  {journal} {\bibinfo  {journal}
  {Phys. Rev. Lett.}\ }\textbf {\bibinfo {volume} {113}},\ \bibinfo {pages}
  {070404} (\bibinfo {year} {2014})}\BibitemShut {NoStop}%
\bibitem [{\citenamefont {Kollár}\ \emph {et~al.}(2015)\citenamefont
  {Kollár}, \citenamefont {Papageorge}, \citenamefont {Baumann}, \citenamefont
  {Armen},\ and\ \citenamefont {Lev}}]{lev_multimode_exp_2015}%
  \BibitemOpen
  \bibfield  {author} {\bibinfo {author} {\bibfnamefont {A.~J.}\ \bibnamefont
  {Kollár}}, \bibinfo {author} {\bibfnamefont {A.~T.}\ \bibnamefont
  {Papageorge}}, \bibinfo {author} {\bibfnamefont {K.}~\bibnamefont {Baumann}},
  \bibinfo {author} {\bibfnamefont {M.~A.}\ \bibnamefont {Armen}}, \ and\
  \bibinfo {author} {\bibfnamefont {B.~L.}\ \bibnamefont {Lev}},\ }\href
  {http://stacks.iop.org/1367-2630/17/i=4/a=043012} {\bibfield  {journal}
  {\bibinfo  {journal} {New Journal of Physics}\ }\textbf {\bibinfo {volume}
  {17}},\ \bibinfo {pages} {043012} (\bibinfo {year} {2015})}\BibitemShut
  {NoStop}%
\bibitem [{\citenamefont {Vetsch}\ \emph {et~al.}(2010)\citenamefont {Vetsch},
  \citenamefont {Reitz}, \citenamefont {Sagu\'e}, \citenamefont {Schmidt},
  \citenamefont {Dawkins},\ and\ \citenamefont
  {Rauschenbeutel}}]{rauschenb_2010}%
  \BibitemOpen
  \bibfield  {author} {\bibinfo {author} {\bibfnamefont {E.}~\bibnamefont
  {Vetsch}}, \bibinfo {author} {\bibfnamefont {D.}~\bibnamefont {Reitz}},
  \bibinfo {author} {\bibfnamefont {G.}~\bibnamefont {Sagu\'e}}, \bibinfo
  {author} {\bibfnamefont {R.}~\bibnamefont {Schmidt}}, \bibinfo {author}
  {\bibfnamefont {S.~T.}\ \bibnamefont {Dawkins}}, \ and\ \bibinfo {author}
  {\bibfnamefont {A.}~\bibnamefont {Rauschenbeutel}},\ }\href {\doibase
  10.1103/PhysRevLett.104.203603} {\bibfield  {journal} {\bibinfo  {journal}
  {Phys. Rev. Lett.}\ }\textbf {\bibinfo {volume} {104}},\ \bibinfo {pages}
  {203603} (\bibinfo {year} {2010})}\BibitemShut {NoStop}%
\bibitem [{\citenamefont {Goban}\ \emph {et~al.}(2012)\citenamefont {Goban},
  \citenamefont {Choi}, \citenamefont {Alton}, \citenamefont {Ding},
  \citenamefont {Lacro\^ute}, \citenamefont {Pototschnig}, \citenamefont
  {Thiele}, \citenamefont {Stern},\ and\ \citenamefont
  {Kimble}}]{kimble_2012_fiber}%
  \BibitemOpen
  \bibfield  {author} {\bibinfo {author} {\bibfnamefont {A.}~\bibnamefont
  {Goban}}, \bibinfo {author} {\bibfnamefont {K.~S.}\ \bibnamefont {Choi}},
  \bibinfo {author} {\bibfnamefont {D.~J.}\ \bibnamefont {Alton}}, \bibinfo
  {author} {\bibfnamefont {D.}~\bibnamefont {Ding}}, \bibinfo {author}
  {\bibfnamefont {C.}~\bibnamefont {Lacro\^ute}}, \bibinfo {author}
  {\bibfnamefont {M.}~\bibnamefont {Pototschnig}}, \bibinfo {author}
  {\bibfnamefont {T.}~\bibnamefont {Thiele}}, \bibinfo {author} {\bibfnamefont
  {N.~P.}\ \bibnamefont {Stern}}, \ and\ \bibinfo {author} {\bibfnamefont
  {H.~J.}\ \bibnamefont {Kimble}},\ }\href {\doibase
  10.1103/PhysRevLett.109.033603} {\bibfield  {journal} {\bibinfo  {journal}
  {Phys. Rev. Lett.}\ }\textbf {\bibinfo {volume} {109}},\ \bibinfo {pages}
  {033603} (\bibinfo {year} {2012})}\BibitemShut {NoStop}%
\bibitem [{\citenamefont {Thompson}\ \emph {et~al.}(2013)\citenamefont
  {Thompson}, \citenamefont {Tiecke}, \citenamefont {de~Leon}, \citenamefont
  {Feist}, \citenamefont {Akimov}, \citenamefont {Gullans}, \citenamefont
  {Zibrov}, \citenamefont {Vuletić},\ and\ \citenamefont
  {Lukin}}]{lukin_2013_fibercavity}%
  \BibitemOpen
  \bibfield  {author} {\bibinfo {author} {\bibfnamefont {J.~D.}\ \bibnamefont
  {Thompson}}, \bibinfo {author} {\bibfnamefont {T.~G.}\ \bibnamefont
  {Tiecke}}, \bibinfo {author} {\bibfnamefont {N.~P.}\ \bibnamefont {de~Leon}},
  \bibinfo {author} {\bibfnamefont {J.}~\bibnamefont {Feist}}, \bibinfo
  {author} {\bibfnamefont {A.~V.}\ \bibnamefont {Akimov}}, \bibinfo {author}
  {\bibfnamefont {M.}~\bibnamefont {Gullans}}, \bibinfo {author} {\bibfnamefont
  {A.~S.}\ \bibnamefont {Zibrov}}, \bibinfo {author} {\bibfnamefont
  {V.}~\bibnamefont {Vuletić}}, \ and\ \bibinfo {author} {\bibfnamefont
  {M.~D.}\ \bibnamefont {Lukin}},\ }\href {\doibase 10.1126/science.1237125}
  {\bibfield  {journal} {\bibinfo  {journal} {Science}\ }\textbf {\bibinfo
  {volume} {340}},\ \bibinfo {pages} {1202} (\bibinfo {year} {2013})},\ \Eprint
  {http://arxiv.org/abs/http://www.sciencemag.org/content/340/6137/1202.full.pdf}
  {http://www.sciencemag.org/content/340/6137/1202.full.pdf} \BibitemShut
  {NoStop}%
\bibitem [{\citenamefont {Goban}\ \emph {et~al.}(2014)\citenamefont {Goban},
  \citenamefont {Hung}, \citenamefont {Yu}, \citenamefont {Hood}, \citenamefont
  {Muniz}, \citenamefont {Lee}, \citenamefont {Martin}, \citenamefont
  {McClung}, \citenamefont {Choi}, \citenamefont {Chang}, \citenamefont
  {Painter},\ and\ \citenamefont {Kimble}}]{kimble_2014_crystal}%
  \BibitemOpen
  \bibfield  {author} {\bibinfo {author} {\bibfnamefont {A.}~\bibnamefont
  {Goban}}, \bibinfo {author} {\bibfnamefont {C.~L.}\ \bibnamefont {Hung}},
  \bibinfo {author} {\bibfnamefont {S.~P.}\ \bibnamefont {Yu}}, \bibinfo
  {author} {\bibfnamefont {J.~D.}\ \bibnamefont {Hood}}, \bibinfo {author}
  {\bibfnamefont {J.~A.}\ \bibnamefont {Muniz}}, \bibinfo {author}
  {\bibfnamefont {J.~H.}\ \bibnamefont {Lee}}, \bibinfo {author} {\bibfnamefont
  {M.~J.}\ \bibnamefont {Martin}}, \bibinfo {author} {\bibfnamefont {A.~C.}\
  \bibnamefont {McClung}}, \bibinfo {author} {\bibfnamefont {K.~S.}\
  \bibnamefont {Choi}}, \bibinfo {author} {\bibfnamefont {D.~E.}\ \bibnamefont
  {Chang}}, \bibinfo {author} {\bibfnamefont {O.}~\bibnamefont {Painter}}, \
  and\ \bibinfo {author} {\bibfnamefont {H.~J.}\ \bibnamefont {Kimble}},\
  }\href {http://dx.doi.org/10.1038/ncomms4808} {\bibfield  {journal} {\bibinfo
   {journal} {Nat Commun}\ }\textbf {\bibinfo {volume} {5}} (\bibinfo {year}
  {2014})}\BibitemShut {NoStop}%
\bibitem [{\citenamefont {Campa}\ \emph {et~al.}(2014)\citenamefont {Campa},
  \citenamefont {Dauxois}, \citenamefont {Fanelli},\ and\ \citenamefont
  {Ruffo}}]{ruffo_review}%
  \BibitemOpen
  \bibfield  {author} {\bibinfo {author} {\bibfnamefont {A.}~\bibnamefont
  {Campa}}, \bibinfo {author} {\bibfnamefont {T.}~\bibnamefont {Dauxois}},
  \bibinfo {author} {\bibfnamefont {D.}~\bibnamefont {Fanelli}}, \ and\
  \bibinfo {author} {\bibfnamefont {S.}~\bibnamefont {Ruffo}},\ }\href@noop {}
  {\emph {\bibinfo {title} {Physics of Long-Range Interacting Systems}}}\
  (\bibinfo  {publisher} {Oxford University Press},\ \bibinfo {year}
  {2014})\BibitemShut {NoStop}%
\bibitem [{\citenamefont {Sch\"utz}\ and\ \citenamefont
  {Morigi}(2014)}]{morigi_pret}%
  \BibitemOpen
  \bibfield  {author} {\bibinfo {author} {\bibfnamefont {S.}~\bibnamefont
  {Sch\"utz}}\ and\ \bibinfo {author} {\bibfnamefont {G.}~\bibnamefont
  {Morigi}},\ }\href {\doibase 10.1103/PhysRevLett.113.203002} {\bibfield
  {journal} {\bibinfo  {journal} {Phys. Rev. Lett.}\ }\textbf {\bibinfo
  {volume} {113}},\ \bibinfo {pages} {203002} (\bibinfo {year}
  {2014})}\BibitemShut {NoStop}%
\bibitem [{\citenamefont {Batrouni}\ \emph {et~al.}(1995)\citenamefont
  {Batrouni}, \citenamefont {Scalettar}, \citenamefont {Zimanyi},\ and\
  \citenamefont {Kampf}}]{batrouni_supersolid_1995}%
  \BibitemOpen
  \bibfield  {author} {\bibinfo {author} {\bibfnamefont {G.~G.}\ \bibnamefont
  {Batrouni}}, \bibinfo {author} {\bibfnamefont {R.~T.}\ \bibnamefont
  {Scalettar}}, \bibinfo {author} {\bibfnamefont {G.~T.}\ \bibnamefont
  {Zimanyi}}, \ and\ \bibinfo {author} {\bibfnamefont {A.~P.}\ \bibnamefont
  {Kampf}},\ }\href {\doibase 10.1103/PhysRevLett.74.2527} {\bibfield
  {journal} {\bibinfo  {journal} {Phys. Rev. Lett.}\ }\textbf {\bibinfo
  {volume} {74}},\ \bibinfo {pages} {2527} (\bibinfo {year}
  {1995})}\BibitemShut {NoStop}%
\bibitem [{\citenamefont {Micheli}, \citenamefont {Brennen},\ and\
  \citenamefont {Zoller}(2006)}]{zoller_topological_2006}%
  \BibitemOpen
  \bibfield  {author} {\bibinfo {author} {\bibfnamefont {A.}~\bibnamefont
  {Micheli}}, \bibinfo {author} {\bibfnamefont {G.~K.}\ \bibnamefont
  {Brennen}}, \ and\ \bibinfo {author} {\bibfnamefont {P.}~\bibnamefont
  {Zoller}},\ }\href {http://dx.doi.org/10.1038/nphys287} {\bibfield  {journal}
  {\bibinfo  {journal} {Nat Phys}\ }\textbf {\bibinfo {volume} {2}},\ \bibinfo
  {pages} {341} (\bibinfo {year} {2006})}\BibitemShut {NoStop}%
\bibitem [{\citenamefont {Domokos}\ and\ \citenamefont
  {Ritsch}(2002)}]{ritsch_2002}%
  \BibitemOpen
  \bibfield  {author} {\bibinfo {author} {\bibfnamefont {P.}~\bibnamefont
  {Domokos}}\ and\ \bibinfo {author} {\bibfnamefont {H.}~\bibnamefont
  {Ritsch}},\ }\href {\doibase 10.1103/PhysRevLett.89.253003} {\bibfield
  {journal} {\bibinfo  {journal} {Phys. Rev. Lett.}\ }\textbf {\bibinfo
  {volume} {89}},\ \bibinfo {pages} {253003} (\bibinfo {year}
  {2002})}\BibitemShut {NoStop}%
\bibitem [{\citenamefont {Ritsch}\ \emph {et~al.}(2013)\citenamefont {Ritsch},
  \citenamefont {Domokos}, \citenamefont {Brennecke},\ and\ \citenamefont
  {Esslinger}}]{cavity_rmp}%
  \BibitemOpen
  \bibfield  {author} {\bibinfo {author} {\bibfnamefont {H.}~\bibnamefont
  {Ritsch}}, \bibinfo {author} {\bibfnamefont {P.}~\bibnamefont {Domokos}},
  \bibinfo {author} {\bibfnamefont {F.}~\bibnamefont {Brennecke}}, \ and\
  \bibinfo {author} {\bibfnamefont {T.}~\bibnamefont {Esslinger}},\ }\href
  {\doibase 10.1103/RevModPhys.85.553} {\bibfield  {journal} {\bibinfo
  {journal} {Rev. Mod. Phys.}\ }\textbf {\bibinfo {volume} {85}},\ \bibinfo
  {pages} {553} (\bibinfo {year} {2013})}\BibitemShut {NoStop}%
\bibitem [{\citenamefont {Chang}, \citenamefont {Cirac},\ and\ \citenamefont
  {Kimble}(2013)}]{chang_2013}%
  \BibitemOpen
  \bibfield  {author} {\bibinfo {author} {\bibfnamefont {D.~E.}\ \bibnamefont
  {Chang}}, \bibinfo {author} {\bibfnamefont {J.~I.}\ \bibnamefont {Cirac}}, \
  and\ \bibinfo {author} {\bibfnamefont {H.~J.}\ \bibnamefont {Kimble}},\
  }\href {\doibase 10.1103/PhysRevLett.110.113606} {\bibfield  {journal}
  {\bibinfo  {journal} {Phys. Rev. Lett.}\ }\textbf {\bibinfo {volume} {110}},\
  \bibinfo {pages} {113606} (\bibinfo {year} {2013})}\BibitemShut {NoStop}%
\bibitem [{\citenamefont {Griesser}\ and\ \citenamefont
  {Ritsch}(2013)}]{griesser_2013}%
  \BibitemOpen
  \bibfield  {author} {\bibinfo {author} {\bibfnamefont {T.}~\bibnamefont
  {Griesser}}\ and\ \bibinfo {author} {\bibfnamefont {H.}~\bibnamefont
  {Ritsch}},\ }\href {\doibase 10.1103/PhysRevLett.111.055702} {\bibfield
  {journal} {\bibinfo  {journal} {Phys. Rev. Lett.}\ }\textbf {\bibinfo
  {volume} {111}},\ \bibinfo {pages} {055702} (\bibinfo {year}
  {2013})}\BibitemShut {NoStop}%
\bibitem [{\citenamefont {Keeling}, \citenamefont {Bhaseen},\ and\
  \citenamefont {Simons}(2014)}]{keeling_2014}%
  \BibitemOpen
  \bibfield  {author} {\bibinfo {author} {\bibfnamefont {J.}~\bibnamefont
  {Keeling}}, \bibinfo {author} {\bibfnamefont {J.}~\bibnamefont {Bhaseen}}, \
  and\ \bibinfo {author} {\bibfnamefont {B.}~\bibnamefont {Simons}},\ }\href
  {\doibase 10.1103/PhysRevLett.112.143002} {\bibfield  {journal} {\bibinfo
  {journal} {Phys. Rev. Lett.}\ }\textbf {\bibinfo {volume} {112}},\ \bibinfo
  {pages} {143002} (\bibinfo {year} {2014})}\BibitemShut {NoStop}%
\bibitem [{\citenamefont {Piazza}\ and\ \citenamefont
  {Strack}(2014)}]{piazza_fermi}%
  \BibitemOpen
  \bibfield  {author} {\bibinfo {author} {\bibfnamefont {F.}~\bibnamefont
  {Piazza}}\ and\ \bibinfo {author} {\bibfnamefont {P.}~\bibnamefont
  {Strack}},\ }\href {\doibase 10.1103/PhysRevLett.112.143003} {\bibfield
  {journal} {\bibinfo  {journal} {Phys. Rev. Lett.}\ }\textbf {\bibinfo
  {volume} {112}},\ \bibinfo {pages} {143003} (\bibinfo {year}
  {2014})}\BibitemShut {NoStop}%
\bibitem [{\citenamefont {Chen}, \citenamefont {Yu},\ and\ \citenamefont
  {Zhai}(2014)}]{zhai_2014}%
  \BibitemOpen
  \bibfield  {author} {\bibinfo {author} {\bibfnamefont {Y.}~\bibnamefont
  {Chen}}, \bibinfo {author} {\bibfnamefont {Z.}~\bibnamefont {Yu}}, \ and\
  \bibinfo {author} {\bibfnamefont {H.}~\bibnamefont {Zhai}},\ }\href {\doibase
  10.1103/PhysRevLett.112.143004} {\bibfield  {journal} {\bibinfo  {journal}
  {Phys. Rev. Lett.}\ }\textbf {\bibinfo {volume} {112}},\ \bibinfo {pages}
  {143004} (\bibinfo {year} {2014})}\BibitemShut {NoStop}%
\bibitem [{\citenamefont {Robb}\ \emph {et~al.}(2015)\citenamefont {Robb},
  \citenamefont {Tesio}, \citenamefont {Oppo}, \citenamefont {Firth},
  \citenamefont {Ackemann},\ and\ \citenamefont
  {Bonifacio}}]{bonifacio_freespace_so_2015}%
  \BibitemOpen
  \bibfield  {author} {\bibinfo {author} {\bibfnamefont {G.~R.~M.}\
  \bibnamefont {Robb}}, \bibinfo {author} {\bibfnamefont {E.}~\bibnamefont
  {Tesio}}, \bibinfo {author} {\bibfnamefont {G.-L.}\ \bibnamefont {Oppo}},
  \bibinfo {author} {\bibfnamefont {W.~J.}\ \bibnamefont {Firth}}, \bibinfo
  {author} {\bibfnamefont {T.}~\bibnamefont {Ackemann}}, \ and\ \bibinfo
  {author} {\bibfnamefont {R.}~\bibnamefont {Bonifacio}},\ }\href {\doibase
  10.1103/PhysRevLett.114.173903} {\bibfield  {journal} {\bibinfo  {journal}
  {Phys. Rev. Lett.}\ }\textbf {\bibinfo {volume} {114}},\ \bibinfo {pages}
  {173903} (\bibinfo {year} {2015})}\BibitemShut {NoStop}%
\bibitem [{\citenamefont {Piazza}, \citenamefont {Strack},\ and\ \citenamefont
  {Zwerger}(2013)}]{piazza_bose}%
  \BibitemOpen
  \bibfield  {author} {\bibinfo {author} {\bibfnamefont {F.}~\bibnamefont
  {Piazza}}, \bibinfo {author} {\bibfnamefont {P.}~\bibnamefont {Strack}}, \
  and\ \bibinfo {author} {\bibfnamefont {W.}~\bibnamefont {Zwerger}},\ }\href
  {\doibase http://dx.doi.org/10.1016/j.aop.2013.08.015} {\bibfield  {journal}
  {\bibinfo  {journal} {Annals of Physics}\ }\textbf {\bibinfo {volume}
  {339}},\ \bibinfo {pages} {135 } (\bibinfo {year} {2013})}\BibitemShut
  {NoStop}%
\bibitem [{\citenamefont {Dicke}(1954)}]{dicke_54}%
  \BibitemOpen
  \bibfield  {author} {\bibinfo {author} {\bibfnamefont {R.~H.}\ \bibnamefont
  {Dicke}},\ }\href {\doibase 10.1103/PhysRev.93.99} {\bibfield  {journal}
  {\bibinfo  {journal} {Phys. Rev.}\ }\textbf {\bibinfo {volume} {93}},\
  \bibinfo {pages} {99} (\bibinfo {year} {1954})}\BibitemShut {NoStop}%
\bibitem [{\citenamefont {Stamper-Kurn}(2014)}]{stamper2014cavity}%
  \BibitemOpen
  \bibfield  {author} {\bibinfo {author} {\bibfnamefont {D.~M.}\ \bibnamefont
  {Stamper-Kurn}},\ }in\ \href@noop {} {\emph {\bibinfo {booktitle} {Cavity
  Optomechanics}}}\ (\bibinfo  {publisher} {Springer},\ \bibinfo {year}
  {2014})\ pp.\ \bibinfo {pages} {283--325}\BibitemShut {NoStop}%
\bibitem [{\citenamefont {Nagy}\ \emph {et~al.}(2010)\citenamefont {Nagy},
  \citenamefont {K\'onya}, \citenamefont {Szirmai},\ and\ \citenamefont
  {Domokos}}]{nagy_2010}%
  \BibitemOpen
  \bibfield  {author} {\bibinfo {author} {\bibfnamefont {D.}~\bibnamefont
  {Nagy}}, \bibinfo {author} {\bibfnamefont {G.}~\bibnamefont {K\'onya}},
  \bibinfo {author} {\bibfnamefont {G.}~\bibnamefont {Szirmai}}, \ and\
  \bibinfo {author} {\bibfnamefont {P.}~\bibnamefont {Domokos}},\ }\href
  {\doibase 10.1103/PhysRevLett.104.130401} {\bibfield  {journal} {\bibinfo
  {journal} {Phys. Rev. Lett.}\ }\textbf {\bibinfo {volume} {104}},\ \bibinfo
  {pages} {130401} (\bibinfo {year} {2010})}\BibitemShut {NoStop}%
\bibitem [{\citenamefont {Mekhov}\ and\ \citenamefont
  {Ritsch}(2012)}]{mekhov2012quantum}%
  \BibitemOpen
  \bibfield  {author} {\bibinfo {author} {\bibfnamefont {I.~B.}\ \bibnamefont
  {Mekhov}}\ and\ \bibinfo {author} {\bibfnamefont {H.}~\bibnamefont
  {Ritsch}},\ }\href@noop {} {\bibfield  {journal} {\bibinfo  {journal}
  {Journal of Physics B: Atomic, Molecular and Optical Physics}\ }\textbf
  {\bibinfo {volume} {45}},\ \bibinfo {pages} {102001} (\bibinfo {year}
  {2012})}\BibitemShut {NoStop}%
\bibitem [{\citenamefont {Carmon}\ \emph {et~al.}(2005)\citenamefont {Carmon},
  \citenamefont {Rokhsari}, \citenamefont {Yang}, \citenamefont {Kippenberg},\
  and\ \citenamefont {Vahala}}]{carmon_opto_exp_2005}%
  \BibitemOpen
  \bibfield  {author} {\bibinfo {author} {\bibfnamefont {T.}~\bibnamefont
  {Carmon}}, \bibinfo {author} {\bibfnamefont {H.}~\bibnamefont {Rokhsari}},
  \bibinfo {author} {\bibfnamefont {L.}~\bibnamefont {Yang}}, \bibinfo {author}
  {\bibfnamefont {T.~J.}\ \bibnamefont {Kippenberg}}, \ and\ \bibinfo {author}
  {\bibfnamefont {K.~J.}\ \bibnamefont {Vahala}},\ }\href {\doibase
  10.1103/PhysRevLett.94.223902} {\bibfield  {journal} {\bibinfo  {journal}
  {Phys. Rev. Lett.}\ }\textbf {\bibinfo {volume} {94}},\ \bibinfo {pages}
  {223902} (\bibinfo {year} {2005})}\BibitemShut {NoStop}%
\bibitem [{\citenamefont {Carmon}, \citenamefont {Cross},\ and\ \citenamefont
  {Vahala}(2007)}]{carmon_opto_chaos_exp_2007}%
  \BibitemOpen
  \bibfield  {author} {\bibinfo {author} {\bibfnamefont {T.}~\bibnamefont
  {Carmon}}, \bibinfo {author} {\bibfnamefont {M.~C.}\ \bibnamefont {Cross}}, \
  and\ \bibinfo {author} {\bibfnamefont {K.~J.}\ \bibnamefont {Vahala}},\
  }\href {\doibase 10.1103/PhysRevLett.98.167203} {\bibfield  {journal}
  {\bibinfo  {journal} {Phys. Rev. Lett.}\ }\textbf {\bibinfo {volume} {98}},\
  \bibinfo {pages} {167203} (\bibinfo {year} {2007})}\BibitemShut {NoStop}%
\bibitem [{\citenamefont {Metzger}\ \emph {et~al.}(2008)\citenamefont
  {Metzger}, \citenamefont {Ludwig}, \citenamefont {Neuenhahn}, \citenamefont
  {Ortlieb}, \citenamefont {Favero}, \citenamefont {Karrai},\ and\
  \citenamefont {Marquardt}}]{marquardt_exp_2008}%
  \BibitemOpen
  \bibfield  {author} {\bibinfo {author} {\bibfnamefont {C.}~\bibnamefont
  {Metzger}}, \bibinfo {author} {\bibfnamefont {M.}~\bibnamefont {Ludwig}},
  \bibinfo {author} {\bibfnamefont {C.}~\bibnamefont {Neuenhahn}}, \bibinfo
  {author} {\bibfnamefont {A.}~\bibnamefont {Ortlieb}}, \bibinfo {author}
  {\bibfnamefont {I.}~\bibnamefont {Favero}}, \bibinfo {author} {\bibfnamefont
  {K.}~\bibnamefont {Karrai}}, \ and\ \bibinfo {author} {\bibfnamefont
  {F.}~\bibnamefont {Marquardt}},\ }\href {\doibase
  10.1103/PhysRevLett.101.133903} {\bibfield  {journal} {\bibinfo  {journal}
  {Phys. Rev. Lett.}\ }\textbf {\bibinfo {volume} {101}},\ \bibinfo {pages}
  {133903} (\bibinfo {year} {2008})}\BibitemShut {NoStop}%
\bibitem [{\citenamefont {Keeling}, \citenamefont {Bhaseen},\ and\
  \citenamefont {Simons}(2010)}]{simons_2010}%
  \BibitemOpen
  \bibfield  {author} {\bibinfo {author} {\bibfnamefont {J.}~\bibnamefont
  {Keeling}}, \bibinfo {author} {\bibfnamefont {M.~J.}\ \bibnamefont
  {Bhaseen}}, \ and\ \bibinfo {author} {\bibfnamefont {B.~D.}\ \bibnamefont
  {Simons}},\ }\href {\doibase 10.1103/PhysRevLett.105.043001} {\bibfield
  {journal} {\bibinfo  {journal} {Phys. Rev. Lett.}\ }\textbf {\bibinfo
  {volume} {105}},\ \bibinfo {pages} {043001} (\bibinfo {year}
  {2010})}\BibitemShut {NoStop}%
\bibitem [{\citenamefont {Emary}\ and\ \citenamefont
  {Brandes}(2003)}]{emary03}%
  \BibitemOpen
  \bibfield  {author} {\bibinfo {author} {\bibfnamefont {C.}~\bibnamefont
  {Emary}}\ and\ \bibinfo {author} {\bibfnamefont {T.}~\bibnamefont
  {Brandes}},\ }\href {\doibase 10.1103/PhysRevE.67.066203} {\bibfield
  {journal} {\bibinfo  {journal} {Phys. Rev. E}\ }\textbf {\bibinfo {volume}
  {67}},\ \bibinfo {pages} {066203} (\bibinfo {year} {2003})}\BibitemShut
  {NoStop}%
\bibitem [{\citenamefont {Altland}\ and\ \citenamefont
  {Haake}(2012)}]{haake_chaos_dicke_2012}%
  \BibitemOpen
  \bibfield  {author} {\bibinfo {author} {\bibfnamefont {A.}~\bibnamefont
  {Altland}}\ and\ \bibinfo {author} {\bibfnamefont {F.}~\bibnamefont
  {Haake}},\ }\href {\doibase 10.1103/PhysRevLett.108.073601} {\bibfield
  {journal} {\bibinfo  {journal} {Phys. Rev. Lett.}\ }\textbf {\bibinfo
  {volume} {108}},\ \bibinfo {pages} {073601} (\bibinfo {year}
  {2012})}\BibitemShut {NoStop}%
\bibitem [{\citenamefont {Bastarrachea-Magnani}\ \emph
  {et~al.}(2015)\citenamefont {Bastarrachea-Magnani}, \citenamefont {del
  Carpio}, \citenamefont {Lerma-Hernández},\ and\ \citenamefont
  {Hirsch}}]{hirsch_dicke_chaos_2015}%
  \BibitemOpen
  \bibfield  {author} {\bibinfo {author} {\bibfnamefont {M.~A.}\ \bibnamefont
  {Bastarrachea-Magnani}}, \bibinfo {author} {\bibfnamefont {B.~L.}\
  \bibnamefont {del Carpio}}, \bibinfo {author} {\bibfnamefont
  {S.}~\bibnamefont {Lerma-Hernández}}, \ and\ \bibinfo {author}
  {\bibfnamefont {J.~G.}\ \bibnamefont {Hirsch}},\ }\href
  {http://stacks.iop.org/1402-4896/90/i=6/a=068015} {\bibfield  {journal}
  {\bibinfo  {journal} {Physica Scripta}\ }\textbf {\bibinfo {volume} {90}},\
  \bibinfo {pages} {068015} (\bibinfo {year} {2015})}\BibitemShut {NoStop}%
\bibitem [{\citenamefont {Brennecke}\ \emph {et~al.}(2008)\citenamefont
  {Brennecke}, \citenamefont {Ritter}, \citenamefont {Donner},\ and\
  \citenamefont {Esslinger}}]{eth_2007}%
  \BibitemOpen
  \bibfield  {author} {\bibinfo {author} {\bibfnamefont {F.}~\bibnamefont
  {Brennecke}}, \bibinfo {author} {\bibfnamefont {S.}~\bibnamefont {Ritter}},
  \bibinfo {author} {\bibfnamefont {T.}~\bibnamefont {Donner}}, \ and\ \bibinfo
  {author} {\bibfnamefont {T.}~\bibnamefont {Esslinger}},\ }\href {\doibase
  10.1126/science.1163218} {\bibfield  {journal} {\bibinfo  {journal}
  {Science}\ }\textbf {\bibinfo {volume} {322}},\ \bibinfo {pages} {235}
  (\bibinfo {year} {2008})}\BibitemShut {NoStop}%
\bibitem [{\citenamefont {Diver}, \citenamefont {Robb},\ and\ \citenamefont
  {Oppo}(2014)}]{oppo_bec_lc_2014}%
  \BibitemOpen
  \bibfield  {author} {\bibinfo {author} {\bibfnamefont {M.}~\bibnamefont
  {Diver}}, \bibinfo {author} {\bibfnamefont {G.~R.~M.}\ \bibnamefont {Robb}},
  \ and\ \bibinfo {author} {\bibfnamefont {G.-L.}\ \bibnamefont {Oppo}},\
  }\href {\doibase 10.1103/PhysRevA.89.033602} {\bibfield  {journal} {\bibinfo
  {journal} {Phys. Rev. A}\ }\textbf {\bibinfo {volume} {89}},\ \bibinfo
  {pages} {033602} (\bibinfo {year} {2014})}\BibitemShut {NoStop}%
\bibitem [{\citenamefont {Ikeda}, \citenamefont {Daido},\ and\ \citenamefont
  {Akimoto}(1980)}]{ikeda_1980}%
  \BibitemOpen
  \bibfield  {author} {\bibinfo {author} {\bibfnamefont {K.}~\bibnamefont
  {Ikeda}}, \bibinfo {author} {\bibfnamefont {H.}~\bibnamefont {Daido}}, \ and\
  \bibinfo {author} {\bibfnamefont {O.}~\bibnamefont {Akimoto}},\ }\href
  {\doibase 10.1103/PhysRevLett.45.709} {\bibfield  {journal} {\bibinfo
  {journal} {Phys. Rev. Lett.}\ }\textbf {\bibinfo {volume} {45}},\ \bibinfo
  {pages} {709} (\bibinfo {year} {1980})}\BibitemShut {NoStop}%
\bibitem [{\citenamefont {Szirmai}, \citenamefont {Mazzarella},\ and\
  \citenamefont {Salasnich}(2015)}]{mazzarella_jos_cav_2015}%
  \BibitemOpen
  \bibfield  {author} {\bibinfo {author} {\bibfnamefont {G.}~\bibnamefont
  {Szirmai}}, \bibinfo {author} {\bibfnamefont {G.}~\bibnamefont {Mazzarella}},
  \ and\ \bibinfo {author} {\bibfnamefont {L.}~\bibnamefont {Salasnich}},\
  }\href {\doibase 10.1103/PhysRevA.91.023601} {\bibfield  {journal} {\bibinfo
  {journal} {Phys. Rev. A}\ }\textbf {\bibinfo {volume} {91}},\ \bibinfo
  {pages} {023601} (\bibinfo {year} {2015})}\BibitemShut {NoStop}%
\bibitem [{\citenamefont {Anderson}(1966)}]{anderson_rmp_1966}%
  \BibitemOpen
  \bibfield  {author} {\bibinfo {author} {\bibfnamefont {P.~W.}\ \bibnamefont
  {Anderson}},\ }\href {\doibase 10.1103/RevModPhys.38.298} {\bibfield
  {journal} {\bibinfo  {journal} {Rev. Mod. Phys.}\ }\textbf {\bibinfo {volume}
  {38}},\ \bibinfo {pages} {298} (\bibinfo {year} {1966})}\BibitemShut
  {NoStop}%
\bibitem [{\citenamefont {Avenel}\ and\ \citenamefont
  {Varoquaux}(1988)}]{avenel_ps_1988}%
  \BibitemOpen
  \bibfield  {author} {\bibinfo {author} {\bibfnamefont {O.}~\bibnamefont
  {Avenel}}\ and\ \bibinfo {author} {\bibfnamefont {E.}~\bibnamefont
  {Varoquaux}},\ }\href {\doibase 10.1103/PhysRevLett.60.416} {\bibfield
  {journal} {\bibinfo  {journal} {Phys. Rev. Lett.}\ }\textbf {\bibinfo
  {volume} {60}},\ \bibinfo {pages} {416} (\bibinfo {year} {1988})}\BibitemShut
  {NoStop}%
\bibitem [{\citenamefont {Amar}\ \emph {et~al.}(1992)\citenamefont {Amar},
  \citenamefont {Sasaki}, \citenamefont {Lozes}, \citenamefont {Davis},\ and\
  \citenamefont {Packard}}]{packard_ps_1992}%
  \BibitemOpen
  \bibfield  {author} {\bibinfo {author} {\bibfnamefont {A.}~\bibnamefont
  {Amar}}, \bibinfo {author} {\bibfnamefont {Y.}~\bibnamefont {Sasaki}},
  \bibinfo {author} {\bibfnamefont {R.~L.}\ \bibnamefont {Lozes}}, \bibinfo
  {author} {\bibfnamefont {J.~C.}\ \bibnamefont {Davis}}, \ and\ \bibinfo
  {author} {\bibfnamefont {R.~E.}\ \bibnamefont {Packard}},\ }\href {\doibase
  10.1103/PhysRevLett.68.2624} {\bibfield  {journal} {\bibinfo  {journal}
  {Phys. Rev. Lett.}\ }\textbf {\bibinfo {volume} {68}},\ \bibinfo {pages}
  {2624} (\bibinfo {year} {1992})}\BibitemShut {NoStop}%
\bibitem [{\citenamefont {Wright}\ \emph {et~al.}(2013)\citenamefont {Wright},
  \citenamefont {Blakestad}, \citenamefont {Lobb}, \citenamefont {Phillips},\
  and\ \citenamefont {Campbell}}]{nist_ps_2013}%
  \BibitemOpen
  \bibfield  {author} {\bibinfo {author} {\bibfnamefont {K.~C.}\ \bibnamefont
  {Wright}}, \bibinfo {author} {\bibfnamefont {R.~B.}\ \bibnamefont
  {Blakestad}}, \bibinfo {author} {\bibfnamefont {C.~J.}\ \bibnamefont {Lobb}},
  \bibinfo {author} {\bibfnamefont {W.~D.}\ \bibnamefont {Phillips}}, \ and\
  \bibinfo {author} {\bibfnamefont {G.~K.}\ \bibnamefont {Campbell}},\ }\href
  {\doibase 10.1103/PhysRevLett.110.025302} {\bibfield  {journal} {\bibinfo
  {journal} {Phys. Rev. Lett.}\ }\textbf {\bibinfo {volume} {110}},\ \bibinfo
  {pages} {025302} (\bibinfo {year} {2013})}\BibitemShut {NoStop}%
\bibitem [{\citenamefont {Stringari}\ and\ \citenamefont
  {Pitaevskii}(2003)}]{string_pit}%
  \BibitemOpen
  \bibfield  {author} {\bibinfo {author} {\bibfnamefont {S.}~\bibnamefont
  {Stringari}}\ and\ \bibinfo {author} {\bibfnamefont {L.}~\bibnamefont
  {Pitaevskii}},\ }\href@noop {} {\emph {\bibinfo {title} {Bose-Einstein
  Condensation}}}\ (\bibinfo  {publisher} {Oxford University Press},\ \bibinfo
  {year} {2003})\BibitemShut {NoStop}%
\bibitem [{\citenamefont {Bloch}, \citenamefont {Dalibard},\ and\ \citenamefont
  {Zwerger}(2008)}]{zwerger_rmp}%
  \BibitemOpen
  \bibfield  {author} {\bibinfo {author} {\bibfnamefont {I.}~\bibnamefont
  {Bloch}}, \bibinfo {author} {\bibfnamefont {J.}~\bibnamefont {Dalibard}}, \
  and\ \bibinfo {author} {\bibfnamefont {W.}~\bibnamefont {Zwerger}},\ }\href
  {\doibase 10.1103/RevModPhys.80.885} {\bibfield  {journal} {\bibinfo
  {journal} {Rev. Mod. Phys.}\ }\textbf {\bibinfo {volume} {80}},\ \bibinfo
  {pages} {885} (\bibinfo {year} {2008})}\BibitemShut {NoStop}%
\bibitem [{\citenamefont {Nagy}, \citenamefont {Szirmai},\ and\ \citenamefont
  {Domokos}(2008)}]{domokos_2008}%
  \BibitemOpen
  \bibfield  {author} {\bibinfo {author} {\bibfnamefont {D.}~\bibnamefont
  {Nagy}}, \bibinfo {author} {\bibfnamefont {G.}~\bibnamefont {Szirmai}}, \
  and\ \bibinfo {author} {\bibfnamefont {P.}~\bibnamefont {Domokos}},\ }\href
  {\doibase 10.1140/epjd/e2008-00074-6} {\bibfield  {journal} {\bibinfo
  {journal} {The European Physical Journal D}\ }\textbf {\bibinfo {volume}
  {48}},\ \bibinfo {pages} {127} (\bibinfo {year} {2008})}\BibitemShut
  {NoStop}%
\bibitem [{\citenamefont {Marwan}\ \emph {et~al.}(2007)\citenamefont {Marwan},
  \citenamefont {Romano}, \citenamefont {Thiel},\ and\ \citenamefont
  {Kurths}}]{marwan_2007}%
  \BibitemOpen
  \bibfield  {author} {\bibinfo {author} {\bibfnamefont {N.}~\bibnamefont
  {Marwan}}, \bibinfo {author} {\bibfnamefont {M.~C.}\ \bibnamefont {Romano}},
  \bibinfo {author} {\bibfnamefont {M.}~\bibnamefont {Thiel}}, \ and\ \bibinfo
  {author} {\bibfnamefont {J.}~\bibnamefont {Kurths}},\ }\href {\doibase
  10.1016/j.physrep.2006.11.001} {\bibfield  {journal} {\bibinfo  {journal}
  {Physics Reports}\ }\textbf {\bibinfo {volume} {438}},\ \bibinfo {pages}
  {237} (\bibinfo {year} {2007})}\BibitemShut {NoStop}%
\bibitem [{\citenamefont {Leggett}(2006)}]{leggett2006quantum}%
  \BibitemOpen
  \bibfield  {author} {\bibinfo {author} {\bibfnamefont {A.~J.}\ \bibnamefont
  {Leggett}},\ }\href@noop {} {\emph {\bibinfo {title} {Quantum liquids: Bose
  condensation and Cooper pairing in condensed-matter systems}}}\ (\bibinfo
  {publisher} {Oxford University Press},\ \bibinfo {year} {2006})\BibitemShut
  {NoStop}%
\bibitem [{\citenamefont {Allen}\ \emph {et~al.}(2014)\citenamefont {Allen},
  \citenamefont {Parker}, \citenamefont {Proukakis},\ and\ \citenamefont
  {Barenghi}}]{barenghi_review_2014}%
  \BibitemOpen
  \bibfield  {author} {\bibinfo {author} {\bibfnamefont {A.~J.}\ \bibnamefont
  {Allen}}, \bibinfo {author} {\bibfnamefont {N.~G.}\ \bibnamefont {Parker}},
  \bibinfo {author} {\bibfnamefont {N.~P.}\ \bibnamefont {Proukakis}}, \ and\
  \bibinfo {author} {\bibfnamefont {C.~F.}\ \bibnamefont {Barenghi}},\ }\href
  {http://stacks.iop.org/1742-6596/544/i=1/a=012023} {\bibfield  {journal}
  {\bibinfo  {journal} {Journal of Physics: Conference Series}\ }\textbf
  {\bibinfo {volume} {544}},\ \bibinfo {pages} {012023} (\bibinfo {year}
  {2014})}\BibitemShut {NoStop}%
\bibitem [{\citenamefont {Sieberer}\ \emph {et~al.}(2014)\citenamefont
  {Sieberer}, \citenamefont {Huber}, \citenamefont {Altman},\ and\
  \citenamefont {Diehl}}]{sieberer_2014}%
  \BibitemOpen
  \bibfield  {author} {\bibinfo {author} {\bibfnamefont {L.~M.}\ \bibnamefont
  {Sieberer}}, \bibinfo {author} {\bibfnamefont {S.~D.}\ \bibnamefont {Huber}},
  \bibinfo {author} {\bibfnamefont {E.}~\bibnamefont {Altman}}, \ and\ \bibinfo
  {author} {\bibfnamefont {S.}~\bibnamefont {Diehl}},\ }\href {\doibase
  10.1103/PhysRevB.89.134310} {\bibfield  {journal} {\bibinfo  {journal} {Phys.
  Rev. B}\ }\textbf {\bibinfo {volume} {89}},\ \bibinfo {pages} {134310}
  (\bibinfo {year} {2014})}\BibitemShut {NoStop}%
\end{thebibliography}%

\end{document}